\documentclass[aps,pra,twocolumn,superscriptaddress]{revtex4-1}
\usepackage{graphicx}
\usepackage{amsmath}
\usepackage{bbold}
\usepackage[T1]{fontenc}
\usepackage{xcolor}
\usepackage{hyperref} 

\newcommand{\beq}{\begin{equation}}
\newcommand{\eeq}{\end{equation}}
\newcommand{\beqs}{\begin{equation*}}
\newcommand{\eeqs}{\end{equation*}}
\newcommand{\beqa}{\begin{eqnarray}}
\newcommand{\eeqa}{\end{eqnarray}}
\newcommand{\beqas}{\begin{eqnarray*}}
\newcommand{\eeqas}{\end{eqnarray*}}
\def\bals#1\eals{\begin{align*}#1\end{align*}}
\def\bal#1\eal{\begin{align}#1\end{align}}

\newcommand{\bcent}{\begin{center}}
\newcommand{\ecent}{\end{center}}

\newcommand{\bitem}{\begin{itemize}}
\newcommand{\eitem}{\end{itemize}}

\makeatletter
\newcommand*\bt{\mathpalette\bt@{.7}}
\newcommand*\bt@[2]{\mathbin{\vcenter{\hbox{\scalebox{#2}{$\m@th#1\bullet$}}}}}
\makeatother

\makeatletter
\newcommand*\ct{\mathpalette\ct@{.7}}
\newcommand*\ct@[2]{\mathbin{\vcenter{\hbox{\scalebox{#2}{$\m@th#1\circ$}}}}}
\makeatother

\begin{document}

\title{Ab initio electronic stationary states for nuclear 
         projectiles in solids}
             
\author{Jessica F. K. Halliday}
\affiliation{Theory of Condensed Matter,
             Cavendish Laboratory, University of Cambridge, 
             J. J. Thomson Ave, Cambridge CB3 0HE, United Kingdom}             
             
\author{Marjan Famili}
\affiliation{Theory of Condensed Matter,
             Cavendish Laboratory, University of Cambridge, 
             J. J. Thomson Ave, Cambridge CB3 0HE, United Kingdom}       
             
\author{Nicol\`o Forcellini}
\affiliation{Beijing Academy of Quantum Information Sciences, 
                Beijing 100193, China }                
             
\author{Emilio Artacho}
\affiliation{Theory of Condensed Matter,
             Cavendish Laboratory, University of Cambridge, 
             J. J. Thomson Ave, Cambridge CB3 0HE, United Kingdom}
\affiliation{CIC Nanogune BRTA and DIPC, Tolosa Hiribidea 76, 
             20018 San Sebastian, Spain}
\affiliation{Ikerbasque, Basque Foundation for Science, 48011 Bilbao, Spain}

\date{\today}

\begin{abstract}
  The process by which a nuclear projectile is decelerated by the 
electrons of the condensed matter it traverses is currently being studied by 
following the explicit dynamics of projectile and electrons from first principles in a 
simulation box with a sample of the host matter in periodic boundary conditions.
  The approach has been quite successful for diverse systems even in the 
strong-coupling regime of maximal dissipation. 
  This technique is here revisited for periodic solids in the light of 
the Floquet theory of stopping, a time-periodic 
scattering framework characterizing the stationary dynamical
solutions for a constant velocity projectile in an infinite solid. 
  The effect of proton projectiles in diamond is studied under that light, 
using time-dependent density-functional theory in real time. 
  The Floquet quasi-energy conserving stationary scattering regime, 
characterized by time-periodic properties such as particle density and 
the time derivative of energy, is obtained for a converged system 
size of one thousand atoms.
  The validity of the customary calculation of electronic stopping power 
from the average slope of the density-functional total energy is discussed.
  Quasi-energy conservation, as well as the implied fundamental 
approximations, are critically reviewed.
\end{abstract}


\maketitle

\section{Introduction}

  The quantum dynamical processes established by a swift nucleus traversing
condensed matter give rise to one of the most canonical problems in 
non-equilibrium quantum physics.
  The projectile slows down when interacting with the degrees of freedom of 
the condensed matter host, which effectively provide a bath for
what is in effect quite a paradigmatic quantum friction problem.
  In addition to its fundamental interest, the problem is of applied 
interest in the contexts of nuclear energy \cite{nuclear},
space radiation \cite{aerospace} and medical applications such as
radiation poisoning and cancer radiotherapy \cite{medical}.

  There are different regimes mostly depending on the charge and the speed 
of the projectile \cite{Sigmund,Sigmund2014}. For a projectile velocity 
$v_p \gtrsim 1$ atomic unit ($1 \; \mathrm{ a.u. } \approx c/ 137$, being $c$ 
the speed of light in vacuum), the projectile energy is transferred mostly to the 
electrons of the host, in what is called an electronic stopping process. 
  The dissipation of projectile kinetic energy induced by a homogeneous
electron liquid (jellium) offers a very appealing model system for
such friction, in some sense a fermion counterpart to the Caldeira-Leggett
quantum friction paradigm \cite{Caldeira1981}.
  This theory was initially approached in the low-velocity
limit \cite{Echenique81, Echenique86, Echenique1990}, and it
proved very successful in the description of the Stokes friction
regime in simple metals.
  It was later extended to finite velocity \cite{Schonhammer1988,Bonig1989,Zaremba1995,Lifschitz1998}, 
or, still in the $v\rightarrow 0$ limit, to non-homogeneous
metals \cite{Nazarov2007}.

  The importance of radiation damage in various technological contexts
has given rise to a demand for estimation of electronic stopping power 
($S_e$, the energy transfer rate from the projectile to the matter electrons) 
for materials beyond simple metals. 
  This need has been addressed by Lindhard's linear response formulation 
for decades \cite{lindhard1954,lindhard1961,lindhard1963,lindhard1965}.
  It is a general approach, allowing,  in principle, for any kind of host matter, 
and amenable to first-principles computations \cite{Nazarov2005,Reining2016}.
It is, however, limited by its fundamental assumption of a weak perturbation.

  Such limitation was overcome with the advent of direct numerical 
simulations in real time, replicating the dynamical process
computationally, where a projectile is placed within a large 
simulation box of the host matter, normally in periodic boundary conditions, 
and dragged across the box at  a given velocity while propagating the 
time-dependent Schr\"odinger equation in discretised real time, monitoring 
the electronic energy, density and wave-functions.  
  This is done both at an empirical tight-binding level
\cite{Mason2007,race2010} and from first-principles using time-dependent
density-functional theory (TDDFT) 
\cite{Pruneda2007,Krasheninnikov2007,Quijada2007, 
Hatcher2008, Correa2012, Zeb2012, Zeb2013, Ojanpera2014, 
Ullah2015,Li2015, Wang2015, Schleife2015, Lim2016,
Quashie2016, Reeves2016, Li2017, Yost2017, Bi2017, 
Ullah2018,Halliday2019}, achieving considerable success in the 
calculation of electronic stopping power for problems clearly beyond 
previous theoretical approaches. 

  Although convergence with respect to technical approximations has 
been thoroughly explored \cite{Kanai2017}, including simulation-box
finite-size convergence, the direct simulation approach rests on
the assumption that the stationary regime expected for an isolated  
constant-velocity projectile is well approximated by the seemingly
stationary situation obtained in the simulations.
  By stationary we mean the regime in which the stopping power 
and deformation density (among other quantities) are time-periodic
as the projectile travels along a space-periodic trajectory in the 
host crystal.

  Such stroboscopically stationary states have been the direct object
of a Floquet theory of electronic stopping \cite{Forcellini2020, Famili2021},
which exploits the underlying discrete translational invariance in
space-time for a single constant-velocity projectile moving along a 
space-periodic trajectory in an infinite crystal.  
  The time-periodic deformation density and stopping power obtained
from the Floquet modes should be reached in the long time limit
(always assuming constant projectile velocity).
  The finite periodic simulation boxes used in the direct approach
cannot aim for the long time limit, however, since after some time,
the projectile re-enters the already excited box from the back, or,
more generally, the effect of periodic replicas of the projectile
start to affect each-other's dynamics, and the simulation is no 
longer describing what intended.
  The aim is therefore for times long enough to establish a 
(stroboscopically) stationary single-projectile regime, but 
short enough that the respective perturbations associated to 
the each of the periodic replicas of the projectile do not have 
the chance to interact.

  It is not known how to determine whether such a regime
is reached for any particular system and velocity.
  The existing studies to date heuristically assess the calculations 
and validate the achievement of that regime by comparing with 
the experimentally obtained electronic stopping power. 
  This work aims for a better assessment and characterization of 
the stroboscopically stationary regime in direct simulations of 
bulk diamond 
irradiated with protons along (100), (110), and (111) high symmetry 
channels, with calculations reaching up to a thousand atoms.
  After assessing the electronic stopping process by validating 
the electronic stopping power, a discussion of the
Floquet theory interpretation of the results is presented, and 
its pertinence is validated by the obtained stationarity of the 
relevant quantities.
  Several key approximations are implied in both the real-time
simulations and the Floquet theory of stopping, such as classical
nuclei, and a constant-velocity projectile. 
  They are well justified in relevant physical regimes, and widely 
used in the community, as already discussed in previous literature.
  Given their relevance to the topic of this paper, a critical review
of the fundamental approximations is presented in the Appendix.
  Finally but importantly, a justification of the customary way of 
obtaining $S_e$ in real-time TDDFT simulations as the average 
slope of the corresponding DFT energy is presented as well in
Section~\ref{sec:validity}.


\section{Method}

\label{sec:Method}

   Electronic excitation characteristics (electronic stopping power, 
particle deformation density and related) for a proton traveling at
constant velocity across bulk diamond are calculated 
using time-dependent density-functional theory \cite{Marques2006}
propagating in discretized real time \cite{Tsolakidis2002}
  The adiabatic local-density approximation (ALDA) is used for the 
exchange-correlation potential, by which the potential at any given time 
is obtained from the particle density at the time (thereby neglecting
memory effects, the non-locality of its time dependence), and it is
also taken in its local-density version (LDA) as parametrized by 
Perdew and Zunger \cite{Perdew1981}. 
  Comparisons with other exchange-correlation functionals have been
performed for other systems elsewhere (see e.g. \cite{Ullah}) and
they are of no consequence to this study.

  A proton is located at an initial position within a simulation box
containing a sample of the material in periodic boundary conditions.
  A conventional DFT calculation is first performed to obtain the 
initial Kohn-Sham wavefunctions. 
  They are then propagated in discretised time using a Crank-Nicolson
integration algorithm \cite{Tsolakidis2002} while the proton is moved at 
a constant velocity through the simulation box.
  The electronic stopping power $S_e$ is then obtaied
as the average slope of the LDA energy as the projectile moves
along its trajectory \cite{Pruneda2007,Krasheninnikov2007,Quijada2007, 
Hatcher2008, Correa2012, Zeb2012, Zeb2013, Ojanpera2014, 
Ullah2015,Li2015, Wang2015, Schleife2015, Lim2016,
Quashie2016, Reeves2016, Li2017, Yost2017, Bi2017, 
Ullah2018,Halliday2019}.

  The TDDFT implementation \cite{Garcia2020} of the {\sc Siesta} code 
is used \cite{Ordejon1996,Artacho1999,Soler2002,Artacho2008}.
  The publicly available open-source version of the program was used as
in its master branch, commit 6c3c0249 of 29 January 2021, accessible
in https://gitlab.com/siesta-project.
  All technical details and approximations (double-$\zeta$ or DZP basis, 
norm-conserving pseudopotentials) are the same as used in
\cite{Halliday2019} which simulated protons through graphite.
  The basis set moving with the atoms (in this case the projectile)
imposes consideration of an evolving basis and Hilbert space
\cite{Artacho2017}.
  The particulars are also as in Ref.~\cite{Halliday2019}.
  The simulation box for the diamond sample is a $5\times 5 \times 5$
supercell of the conventional cubic diamond cell of 8 atoms, amounting
to a cubic box with 1000 C atoms (plus one H projectile).
  The lattice parameter used was 3.567 \AA.

\begin{figure}[t]
\centering
\includegraphics[width=0.23\textwidth]{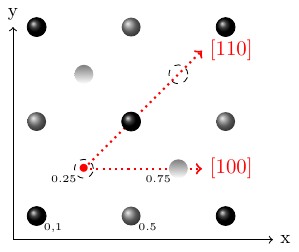}
\includegraphics[width=0.23\textwidth]{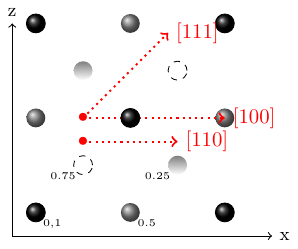}
\caption{(a) Diamond unit cell, looking down the [001] direction, 
showing the initial position of the projectile and the trajectory 
taken along the [100] and [110] directions. 
  (b) Diamond unit cell, looking along the [010] direction, 
showing the initial heights on the projectiles along the $z$ 
axis and the initial trajectories.}
\label{fig:structure-sketch}
\end{figure}


\section{Results and discussion}

\subsection{Stopping power for protons in diamond}  

\label{sec:Results-diamond}
  
\begin{figure}[t]
\centering
\includegraphics[width=0.45\textwidth]{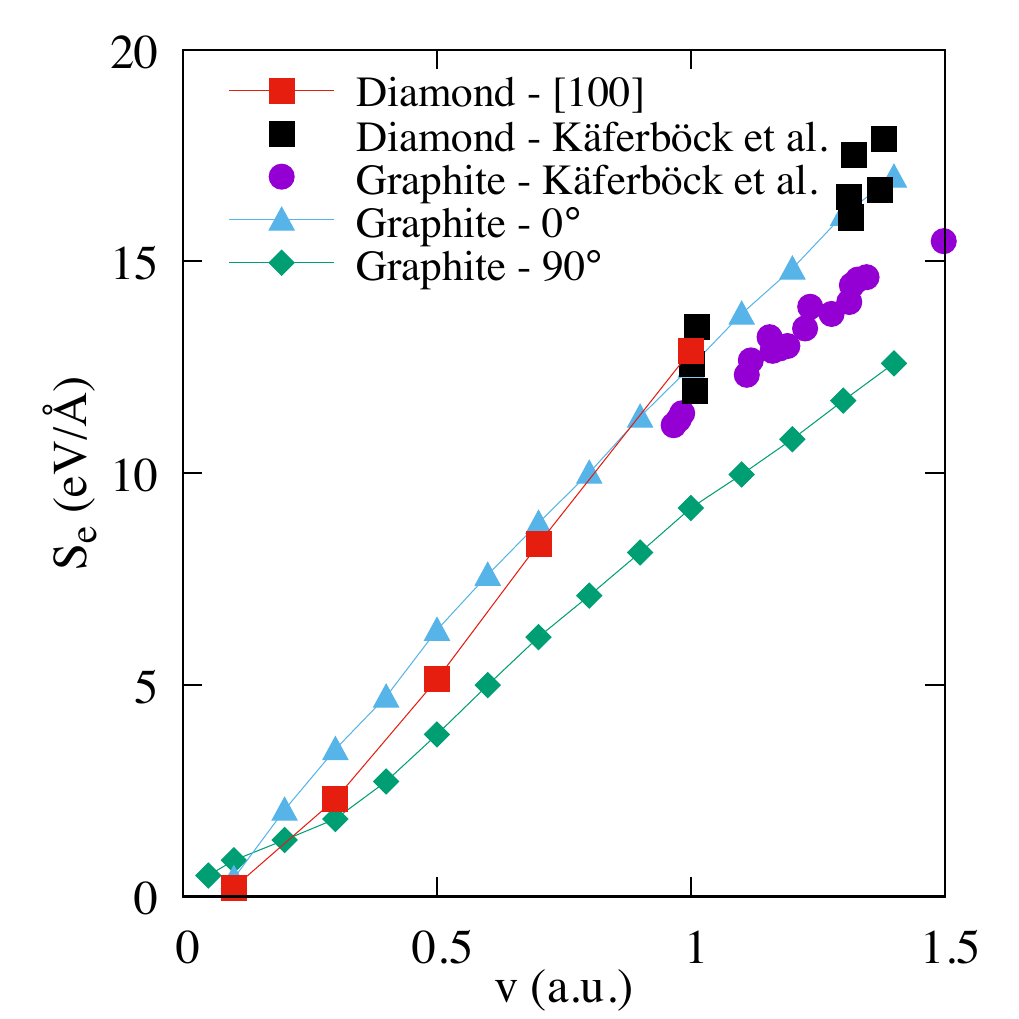}
\caption{Electronic stopping power $S_e$ versus proton velocity
in diamond and graphite. Simulation results for the [100] channelling 
trajectory in diamond (red squares) are compared  with the 
experimental results of Ref.~\cite{kaferbock} for diamond (black squares)
alongside a similar comparison for simulation results in graphite
(parallel and perpendicular to graphitic planes, triangles and
rhombi, respectively) from Ref.~\cite{Halliday2019}, and 
corresponding experimental values from the same experimental
paper ~\cite{kaferbock} (circles).}
\label{fig:diamond-graphite}
\end{figure}

\begin{figure}[t]
\centering
\includegraphics[width=0.45\textwidth]{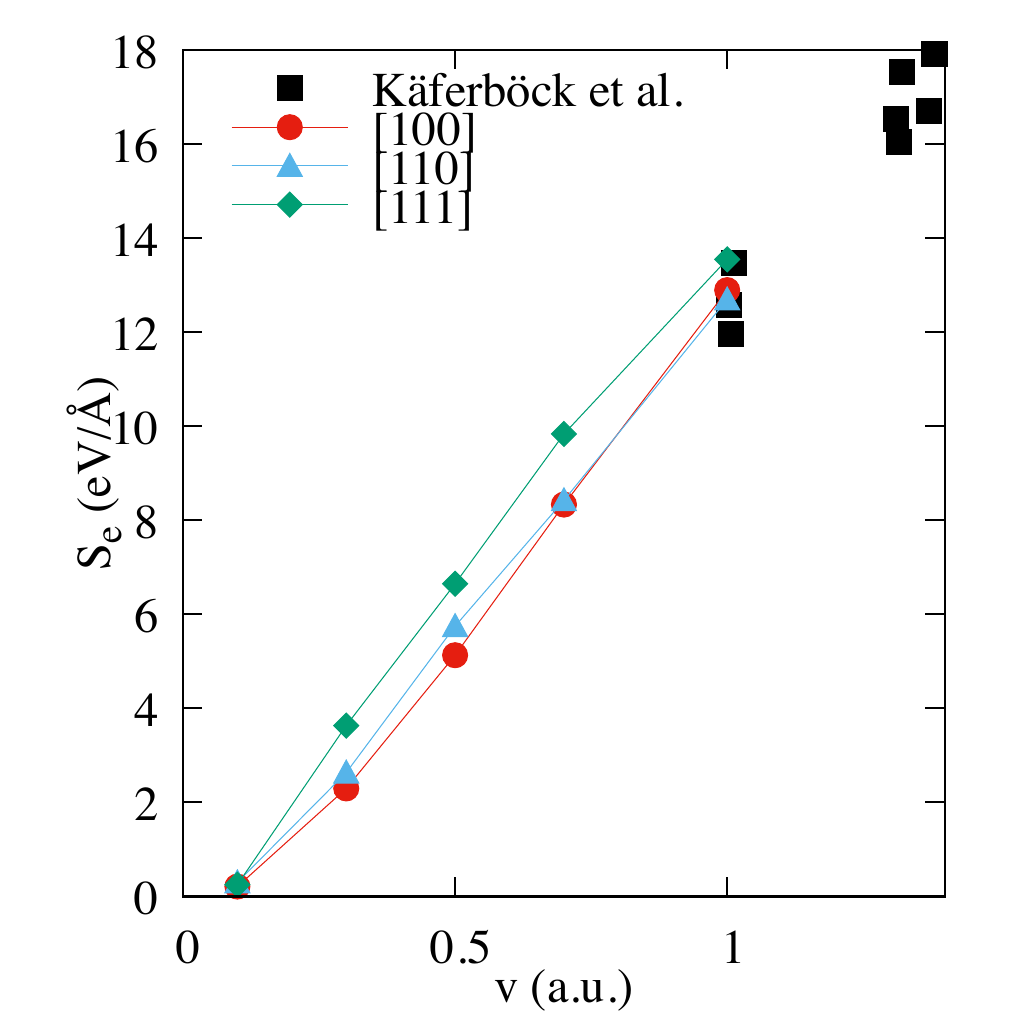}
\caption{Electronic stopping power $S_e$ versus proton velocity
in diamond. Simulation results for channelling trajectories along 
[100], [110], and [111]  directions (circles, triangles and rhombi,
respectively), compared with the experimental
results of Ref.~\cite{kaferbock} (squares).}
\label{fig:diamond-stopping}
\end{figure}

  Figures~\ref{fig:diamond-graphite} and 
\ref{fig:diamond-stopping} show the results of 
the calculation of the electronic stopping power $S_e$ for
protons in diamond as a function of the projectile velocity,
obtained as explained in the Method Section, essentially 
following Refs.~\cite{Pruneda2007,Halliday2019}.
  In this work we concentrate on velocities below the Bragg 
peak, a velocity range already well into the non-adiabatic
regime, but still sensitive to the electronic structure of
the host material (Bragg peak in the sense of the maximum in the 
$S_e(v)$ curve, instead of the conventional meaning
of the maximum of $S_e$ versus penetration; we will use
the former meaning henceforth).
  This is apparent in Fig.~\ref{fig:diamond-graphite},
where the calculated $S_e$ for diamond (red squares) is compared 
with the results for graphite of Ref.~\cite{Halliday2019} (blue and green) 
and both are compared with the respective experimental results of 
Ref.~\cite{kaferbock} (black and purple).
  The anisotropy in graphite gives rise to a wide band of 
theoretical values, already discussed in Ref.~\cite{Halliday2019}.
  However, they still allow to replicate the experimentally observed 
tendency towards larger stopping power for diamond for 
$v_p\gtrsim 1$ a.u.

  Furthermore, the theoretical results display a crossover of the
stopping powers where $S_e^{\mathrm{diamond}} < 
S_e^{\mathrm{graphite}}$ at lower velocities.
  This is a result of the threshold behavior, expected and 
apparent for diamond due to its large band gap.
  This fact cannot be validated with the available 
experimental data, but it does seem plausible given
what is known for other large-band-gap insulators
\cite{Bauer2005,Pruneda2007,Bauer2009,Bauer2017}.
  The anisotropy in $S_e$ in diamond is less pronounced 
than in graphite, as shown in Fig.~\ref{fig:diamond-stopping}, 
which is again expected,
in spite of the fact that the channels considered are 
quite different (in e.g. electronic density).
  Detailed linear-response analysis on the variability of 
$S_e$ for different allotropes of C and different trajectories 
can be found in Ref.~\cite{GarciaMolina2006}.
  This work is about stationary states along channels, 
and what shown in Figs.~\ref{fig:diamond-graphite} and 
\ref{fig:diamond-stopping} is sufficient for supporting
the analysis and conclusions of this paper.

\begin{figure}[t]
\centering
\includegraphics[width=0.4\textwidth]{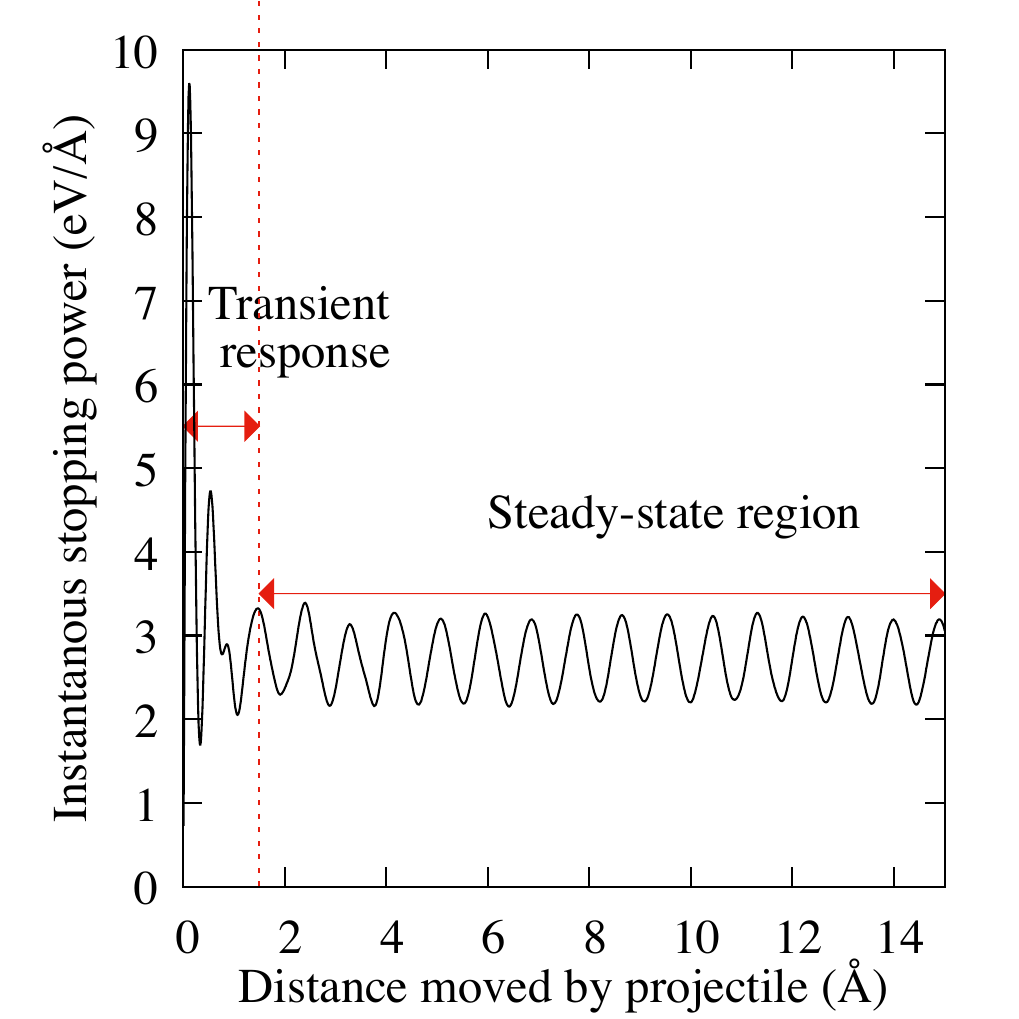}
\caption{Instantaneous electronic stopping power in diamond,  
for a proton moving through the simulation box at $v_p=0.5$ a.u. 
along the [100] direction.}
\label{fig:instantaneous-stopping1}
\end{figure}

  The results of Figs.~\ref{fig:diamond-graphite} and 
\ref{fig:diamond-stopping} are obtained by performing simulations
at fixed $v_p$, and obtaining the instantaneous 
stopping power from 
\beq
\label{eq:inst-stopping}
S_e(t)=\partial_t E_{\mathrm{ALDA}}(t)/v_p \, ,
\eeq
as displayed in Fig.~\ref{fig:instantaneous-stopping1}
versus projectile position, for a particular $v_p$ value.
  The $S_e(t)$ profile in the figure includes an initial interval 
showing a transient response to the abrupt start of the projectile, 
which largely disappears within the first 2 \AA, except for a smaller, 
more persistent oscillating transient. 
  After the initial response, the steady state is quickly established.
 The steady state oscillation gives a constant enough average 
to allow extracting the $S_e$ values for each projectile velocity, 
as are displayed in Figs.~\ref{fig:diamond-graphite} and 
\ref{fig:diamond-stopping}.

  The range of velocities for which calculations have been 
performed is the one optimizing the likelihood to obtain a good
characterization of a steady state for the single-projectile problem, 
which is the aim of this study.
  For larger velocities the projectile traverses the cell in too short a
time for the transient to disappear.
  We have gone up in velocity for our results to connect with 
experimental values, for validation, but not beyond (it respresents
a substantial computational effort). 
  In the low velocity limit the perturbation has time enough to 
propagate among periodic replicas, towards saturation 
(see Section~\ref{sec:saturation}).

\subsection{Validity of  $S_e$ from $\partial_t E_{\mathrm{ADFT}}$ }
\label{sec:validity}

  Extracting the electronic stopping power $S_e$ from
Eq.~\eqref{eq:inst-stopping} as done in the previous subsection,
has been routinely done in TDDFT simulations of stopping using 
adiabatic forms of the exchange and correlation functional (ADFT) 
\cite{Pruneda2007,Krasheninnikov2007,Quijada2007, 
Hatcher2008, Correa2012, Zeb2012, Zeb2013, Ojanpera2014, 
Ullah2015,Li2015, Wang2015, Schleife2015, Lim2016,
Quashie2016, Reeves2016, Li2017, Yost2017, Bi2017, 
Ullah2018,Halliday2019},
since their beginning \cite{Pruneda2007}.
  It is a sensible approach, which has been supported by agreement
with experiments and by a conservation argument \cite{Correa-Review}.
  To our knowledge, however, it has not been more formally justified
so far, which we do here.

  It is indeed quite a different approach to the ones in the pre-existing
literature, most prominently, the one based on scattering 
theory in jellium \cite{Echenique81} and its
generalisations \cite{Schonhammer1988,Bonig1989,Zaremba1995,
Lifschitz1998,Forcellini2020}.
  There, $S_e$ is given essentially by stating the energy transfer rate for 
each incoming state $i$ (occupied in the lab frame) as 
\beq
\label{eq:trad-erate}
\partial_t E_i = n_i \sum_o (\epsilon_o-\epsilon_i) | 
\mathcal{S}_{i\rightarrow o}|^2 v_o
\eeq
where $\epsilon_i$ and $\epsilon_o$ are the respective single-particle
energies of incoming and outgoing waves in the lab frame,
$\mathcal{S}_{i\rightarrow o}$ is the scattering amplitude for that
process, $v_o$ the outgoing group velocity, and $n_i$ the incoming density,
all now in the projectile frame.
  The stopping power then follows summing over all scattering processes
from all the incoming occupied states to compatible unoccupied states 
in the lab frame divided by the projectile velocity for conversion,  as 
$S_e\sim (\sum_i \partial_t E_i)/ v_p$.
  
  Although perfectly sound and valid for independent electrons,
such an expression for $S_e$ within a DFT framework relies on the 
relation between the total electronic energy and the single-particle 
Kohn-Sham eigenvalues.
  That connection is made through Janak's theorem \cite{Janak}, 
but, as it is well known in the context of the insulator band-gap problem
\cite{Sham1983}, it implies a further approximation, especially for
non metals.
  This is elegantly shown in Ref.~\cite{Nazarov2005}, where 
a correct expression is presented in the same vein using
TDDFT in the frequency domain for the low-projectile-velocity 
limit in metals.
  A  further many-electron term is there shown to appear for 
the correct stopping power expression for a given XC functional.

  The $S_e$ calculation in real-time TDDFT simulations 
as extracted from Eq.~\eqref{eq:inst-stopping} is not limited to 
low-velocity projectiles or metals. 
  Its justification can be seen as follows.
  TDDFT with the exact XC (and with proper account of
the initial state) would give the exact evolving particle density
$n(\mathbf{r},t)$, in addition to the relevant action 
\cite{Runge,Marques2006}.
  TDDFT also states that any physical property would be 
expressible as a functional  of that density, although,
a priori, the expression of that functional would not be known
in general.

  However, the functional for the electronic stopping power
is known.
  It is nothing but the net force on the projectile 
\cite{Grande2016,Correa-Review}, 
suitably averaged along the trajectory (and over 
different trajectories depending on the experimental
conditions). 
  As  long as the potential $V_{ext}$ describing the interaction 
between each electron and the nuclei is local, which is a pre-requisite 
for TDDFT (and DFT in general), the instantaneous force on 
the projectile due to the electrons is simply 
\beq
\label{eq:force}
\mathbf{F}_p(t) = - \int \! \! \mathrm{d}^3 \mathbf{r} \, \,
n(\mathbf{r},t) \, \, \nabla_p  V_{ext}  \; ,
\eeq
where $V_{ext}$ is the single-particle external potential
acting on electrons, due to nuclear attractions and other
possible external fields.
  This expression is obtained very generally for Ehrenfest
dynamics using Newton's law for the classical evolution of the
nuclei and TDDFT for the quantum evolution of
electrons \cite{Schmidt1996,Todorov2001,Tavernelli-Review}.
  In our case, for a classical nuclear projectile of constant
velocity and other classical nuclei at rest, it is easy to 
corroborate, given the evolution of the electronic energy
\bal
\label{eq:energyderivative}
\partial_t E_e &= \partial_t \langle \Psi|H|\Psi\rangle
= \langle \Psi|\partial_t H|\Psi\rangle
= \mathbf{v}_p \cdot \langle \Psi|\nabla_p H|\Psi\rangle = \nonumber \\
&=  \mathbf{v}_p \cdot \int \! \! \mathrm{d}^3 \mathbf{r} \, \,
n(\mathbf{r},t) \, \, \nabla_p  V_{ext} = - \mathbf{v}_p \cdot
\mathbf{F}_p
\eal
for any $\mathbf{v}_p$, and where we have used $|\Psi\rangle$
for the exact evolving electronic state and $H$ for the 
electronic Hamiltonian,  which fulfil  
$H |\Psi\rangle = i\hbar \partial_t |\Psi\rangle$,
and the fact that the only explicit time-dependence in 
the electronic Hamiltonian is given by the motion of
the projectile.
  Therefore, a trajectory average of the force in Eq.~\eqref{eq:force} 
gives the electronic stopping power $S_e$ within TDDFT and 
the approximations described so far.

  Most of the electronic stopping power evaluations in direct
real-time TDDFT simulations resort to the electronic energy directly
\cite{Pruneda2007,Correa-Review}, by extracting
$S_e$ from the average slope of the electronic energy 
as the projectile progresses, as in Eq.~\eqref{eq:inst-stopping}.
  For the adiabatic functionals they use, the electronic 
energy is what given by the corresponding
time-independent density functional for the instantaneous 
density, in this case LDA.
  In that sense, $E_{\mathrm{ADFT}} = 
E_{\mathrm{LDA}}[n(\mathbf{r},t)]$ is taken as the
functional of the density for $\langle \Psi (t) | H (t) |\Psi(t) \rangle$,
using the nomenclature of Eq.~\eqref{eq:energyderivative}.
  Firstly, it leads to the consistently correct adiabatic energy in the 
$v_p \rightarrow 0$ limit.
  Secondly, for any adiabatic time-dependent XC functional
the XC action can be expressed as 
\beqs
A_{xc}[n(\mathbf{r},t)] = \int \mathrm{d} t \, E_{xc}[n(\mathbf{r})](t)
\eeqs
and then, the quantity
\beqs
E_{tot} = E_{\mathrm{ADFT}}+T_N+\sum_{I,J<I} \frac{Z_I Z_J}{R_{IJ}}  
\eeqs
is shown to be conserved in Ehrenfest dynamics 
\cite{Schmidt1996,Todorov2001,Kunert2003}, where
$T_N$ stands for the nuclear kinetic energy.
  From Eq.~\eqref{eq:energyderivative}, it can be seen that
the calculation of $S_e$  from  the slope of the electronic energy 
and from the force as in Eq.~\eqref{eq:force} are equivalent.
  Therefore, a calculation of $S_e$ based on
Eq.~\eqref{eq:inst-stopping} is well supported by TDDFT,
and gives the correct value within the theory defined by
the chosen XC, as long as it is an adiabatic functional.
  If the non-locality in time is considered, Eq.~\eqref{eq:inst-stopping}
should not be used, but rather $S_e$ should be extracted 
from the average force on the projectile, Eq.~\eqref{eq:force}.


\subsection{Steady state}

\begin{figure}[t]
\centering
\includegraphics[width=0.4\textwidth]{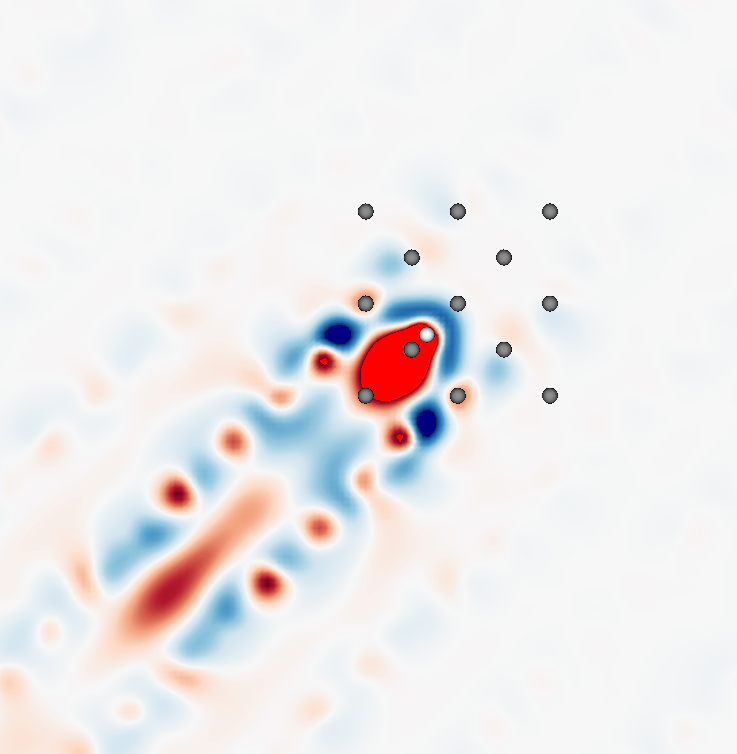}
\vskip 6pt
\includegraphics[width=0.4\textwidth]{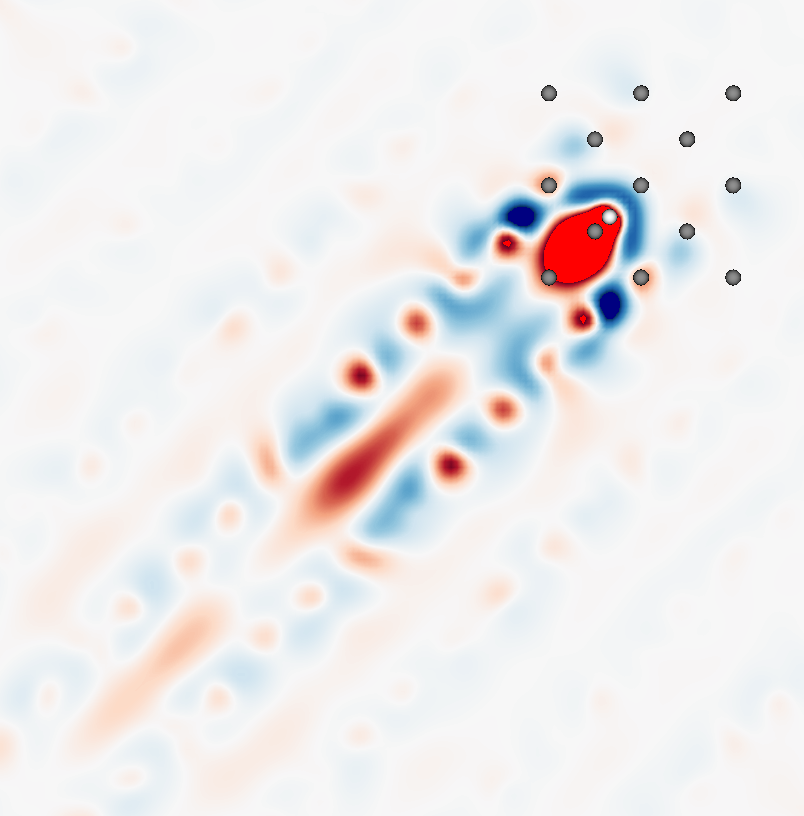}
\caption{Electron deformation density in real space at equivalent 
projectile positions in consecutive diamond unit cells (upper and lower
panels) along the trajectory for a proton moving at $v=1$ a.u. along 
the [110] direction. 
  It is depicted in the (001) plane containing the projectile,
the horizontal axis being the [100] direction and the vertical 
axis the [010] direction.
  The color scale is given from $-0.01e/$Bohr (dark blue) to
$+0.01e/$Bohr (dark red), going through white for zero.
  The homogeneous bright red close to the proton is
saturated (scale chosen to show rich pattern in the tail).
Beads indicate atomic positions of selected atoms including
the projectile (light bead) to indicate the equivalence of 
position under translation.}
\label{fig:deformation-density}
\end{figure}

  The Floquet theory of electronic stopping \cite{Forcellini2020,Famili2021}
describes the stationary state for a single projectile at constant velocity
the approximations and assumptions implied, such as a non-relativistic
classical nuclear projectile and constant velocity, are critically reviewed
in the Appendix, while size effects in the approach to an apparent 
stationary state are reviewed in \cite{Correa-Review}).  
  It is a stroboscopically stationary state, meaning that properties that do 
not scale with the size of the system are invariant when looked at times 
separated  by a time period $\tau = a/v_p$, where $a$ is the lattice parameter of 
the one-dimensional lattice along the rectilinear projectile trajectory.
  The particle density $n(\mathbf{r},t)$, the force opposing the projectile
motion $\mathbf{F}_p(t)$, and, therefore, the electron energy excitation rate, 
$\partial_t E_e = v_p S_e$,  are all stroboscopically stationary quantitites, 
both in an exact solution and within our ALDA approach. 

  Since the time average of $\partial_t E_e$ is not zero, the electronic 
energy itself steadily grows, and it is therefore not a stroboscopically 
stationary property.
  This is compatible with Floquet theorem since it is an infinite open system.
  It is analogous to a scattering problem in general, and to
the stopping problem in jellium, in particular, which is time independent, 
with constant $S_e$ and therefore increasing $E_e$. 
  There are properties relating to the Kohn-Sham single-particle states
that are also stroboscopically stationary
\footnote{See e.g. 
the spectral projections presented in Ref.~\cite{Zeb2012}.
  Scattering amplitudes can be defined for the single-particle 
scattering processes in the projectile reference frame
and within the Floquet theory \cite{Forcellini2020,Famili2021}.
  Matrix elements that make sense for 
supercell time-dependent simulations
may, however, steadily increase with time due to the flux implied
in scattering (see Fig.~2 in Ref.~\cite{Zeb2012}).}, but
we concentrate here on energy, density and forces.

\begin{figure}[t]
\centering
\includegraphics[width=0.47\textwidth]{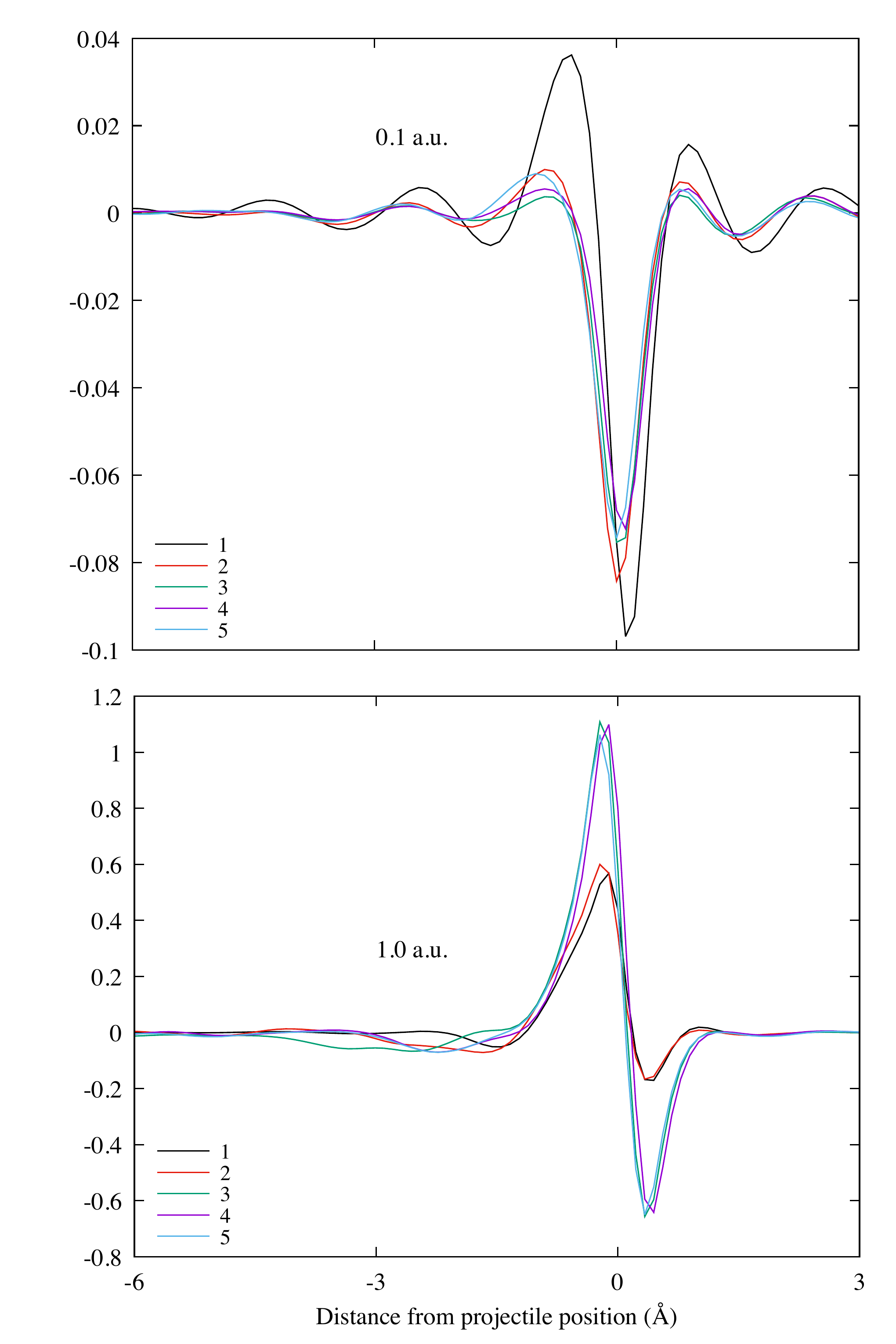}
\caption{Electron deformation density along the path of the projectile, 
comparing for projectile positions at five different points in the 
simulation box which are crystallographically equivalent, 
labelled 1-5, the number increasing the more advanced the projectile.
Upper (lower) panel is for $v_p=0.1$ a.u. (1.0 a.u.).}
\label{fig:tail-deformation-density}
\end{figure}

  Figure~\ref{fig:instantaneous-stopping1} clearly displays 
the described expectation for $\partial_t E_e$: after the 
transient, an oscillatory behavior for, in this case, the 
time derivative of the electronic energy, with the expected 
period (the lattice parameter along [100] is $a=3.567$ \AA,
the projectile passing close to a C atom every $a/4 = 0.89$ \AA,
which perfectly conforms with the observed periodicity). 
  Fig.~\ref{fig:deformation-density} shows the repetition of
the dynamical deformation density at equivalent proton positions 
(and corresponding times) in the diamond crystal, for the projectile 
moving at $v_p=1$ a.u. along the [110] direction.
  The dynamical deformation density is defined as
\beqs
\delta n^{\mathrm{dyn}}(\mathbf{r},t) = n(\mathbf{r},t) - 
n^a[\mathbf{r},\{\mathbf{R}_I(t)\} ]
\eeqs
$n(\mathbf{r},t)$ being the dynamical electron density, and
$n^a[\mathbf{r},\{\mathbf{R}_I(t)\} ]$ the corresponding adiabatic
electron density, meaning the one corresponding to the
electronic ground state for the nuclei at their $\mathbf{R}_I$ positions 
at time $t$.

  There are depictions in the literature of the wake of the 
projectile in terms of maps of the density or the deformation
density (see e.g. very early sets in Refs.~\cite{Ritchie1976,Echenique1979} 
for jellium, and also for metals such as in the cover of the issue for 
Ref.~\cite{Correa2012} and in Refs.~\cite{Quashie2016,Correa-Review}
for example). 
  A comparative wake analysis for different
velocities, directions, and materials would be of interest.
  It is, however, beyond the scope of this work: our focus is on
the stroboscopic stationarity of the density as key magnitude
in first-principles calculations of these processes. 
  The key comparison is therefore between the upper and lower
panels of Fig.~\ref{fig:deformation-density}.
  Indeed, the deformation density pattern is very rich and intricate,
and the  similarity under translation is remarkable.

  For a more quantitative assessment, Fig.~\ref{fig:tail-deformation-density}
shows the same deformation density along the projectile path 
at different projectile positions that are crystallographically 
equivalent, taking the projectile position as reference for 
better comparison. 
  It is shown for low (0.1 a.u.) and high (1.0 a.u.) velocity. 
  It can be seen how the deformation tends to stabilise in 
the latter positions, once the projectile has approached the 
stationary state.
  It can be appreciated that the stationarity is only approximate, 
with small differences between the last two points, points 4 and 5 
(spiky features in the lower panel relate to the real-space 
discretisation inherent to the SIESTA method \cite{Soler2002}).

\subsection{Saturation}
\label{sec:saturation}

\begin{figure}[t]
\centering
\includegraphics[width=0.4\textwidth]{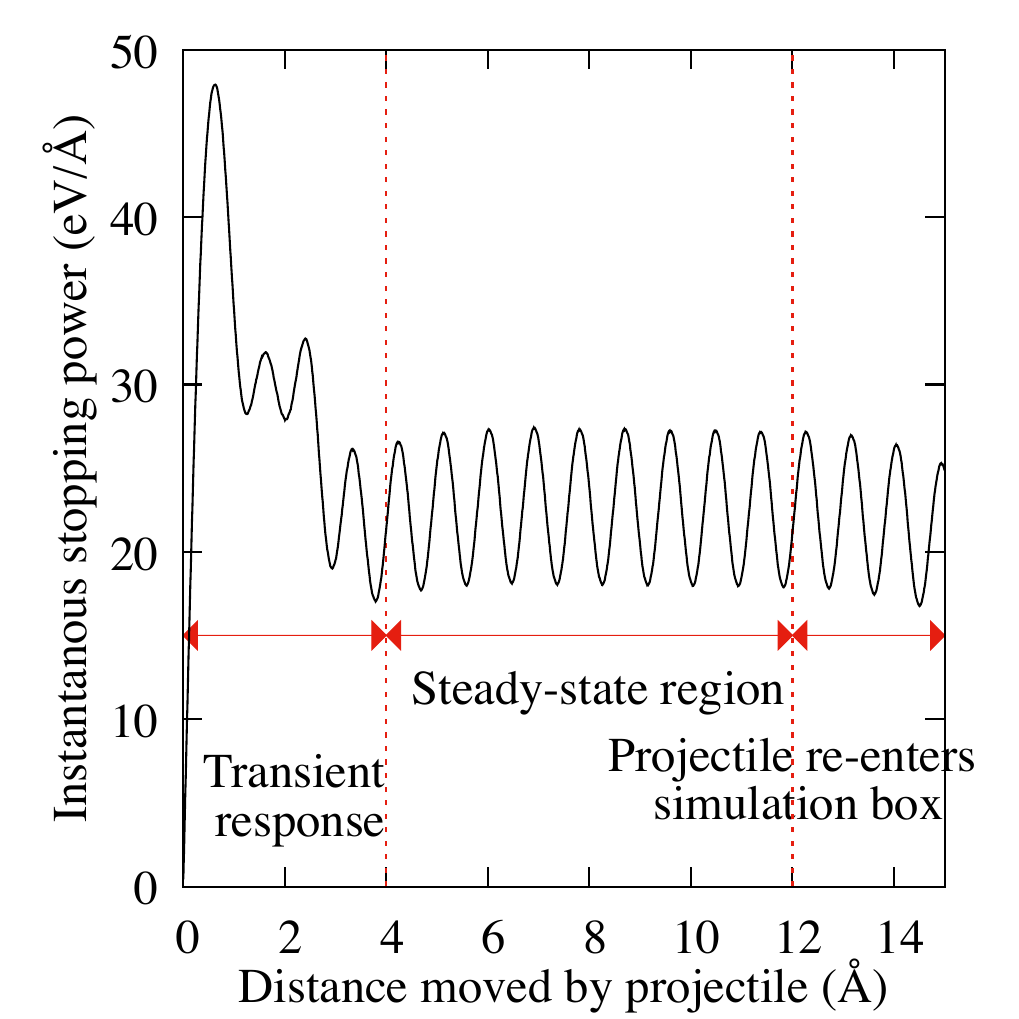}
\caption{Instantaneous electronic stopping power in diamond for
a proton moving at $v_p=0.1$ a.u. along the [100] direction.}
\label{fig:instantaneous-stopping2}
\end{figure}

  For the single-projectile problem, the Floquet stationary solutions 
would be reached in the long-time limit (if at all).
  It should be noted, however, that in the periodically repeated 
projectile simulations described in Sections~\ref{sec:Method} and 
\ref{sec:Results-diamond}, in the long-time limit, a different regime 
is to be expected. 
  It is also a periodic problem, but one in which, due to its 
effective finite size (or finite host size per projectile), Floquet modes 
would be expected for which the electronic energy itself (not its flux)
would be periodic, and, therefore, the oscillating $S_e(t)$ would 
average to zero.
  Physically, since projectile replicas pump energy into the host 
electronic system everywhere, the system should reach a 
saturated periodic state, as described for finite Floquet 
systems (see for instance \cite{Alessio2014,lazarides2014}), in which 
energy would be transferred back and forth between the projectile lattice 
and the electrons. 
  It is not a regime of interest to the single-projectile problem, 
but can be rather considered as a finite-size effect, and the 
simulation box size is normally increased when hints of 
saturation appear before a meaningful description of the 
intended phenomena is reached.
  Saturation will not be characterised here, beyond illustrating its 
onset in Fig.~\ref{fig:instantaneous-stopping2}.

  The calculation of $S_e$ and characterisation of processes 
of electronic stopping of projectiles by means of direct 
computation in a finite simulation box relies on the system 
reaching a regime in which the effect of other projectile 
replicas is still not felt by a given projectile (finding a ``fresh'' 
host as it moves).
  Each replica hence approximately behaves as a single 
projectile, and, therefore, could achieve the stationary 
regime predicted by the Floquet theory of stopping 
\cite{Forcellini2020}, before the effect of other projectiles 
is felt (e.g. by re-entering the cell, or energy irradiation from
other replicas reaching the one being followed).

  The sweet spot is therefore in the interval between the end 
of the initial transient and the onset of saturation, as illustrated 
in Fig.~\ref{fig:instantaneous-stopping2}.
  It is so far hard to foretell the duration, and even the existence, 
of such a regime.
  It will depend on the projectile velocity and kind of host, 
as well as the type of start (smooth vs instantaneous).
  So far, it is addressed as a technical system-size 
convergence problem (see a discussion on this in 
\cite{Schleife2015}), and, so far, this community 
has been fortunate in finding most situations amenable 
to feasible computation.

  A more fundamental problem, however, is, once a sta-ble 
oscillation is found, to ascertain whether what observed really 
corresponds to the actually sought Floquet stationary regime 
of the stopping problem.
  The ``thermalisation'' of such a system into the steady 
non-equilibrium regime does not appear to be a trivial problem.
  Some correlated systems are known not to thermalise 
to equilibrium while stuck in a many-body localised state 
\cite{abanin2019,ponte2015,lazarides2015}. 
  In the strongly out of equilibrium problem presented by 
electronic stopping in various kinds of solids we are not 
aware of any way of ascertaining on reaching the relevant 
steady state.
  Helpful here would be the calculation of the Floquet 
stationary state directly on a TDDFT implementation of 
the scattering Floquet theory of stopping, analogous 
to what accomplished in Ref.~\cite{Famili2021} for a 
tight-binding Hamiltonian. 
  That should be the focus of further work.


\section{Conclusions}

  Based on results for the electronic stopping power for
protons in diamond, the results of such simulations are 
analyzed in the light of the stroboscopic stationary states 
predicted by the recent Floquet theory of stopping. 
  The predicted Floquet solutions for the electronic 
excitation by a projectile are distinguished from the 
ones expected for a lattice of projectiles, the latter
corresponding to the actual supercell calculations, 
the former to the physical process of interest.

  The intermediate regime between the initial transient
behavior due to the start of the projectile
motion and the onset of saturation is shown
to reflect the stroboscopic stationary state of the 
single projectile, and the stationary character
is tested quantitatively.
  The results for $S_e$ and particle density for protons 
in diamond display remarkable stroboscopic invariance 
of both magnitudes in that regime.
  
  The average of $\partial_t E_e/v_p$ is 
shown to give the correct electronic stopping power $S_e$,
within TDDFT, supporting the way it has been calculated in many
electronic stopping studies, but only for adiabatic
time-dependent exchange-correlation functionals.
  For density functionals beyond that approximation 
(contemplating the non-locality of the time dependence),
the force on the projectile should be used instead.

\begin{acknowledgments}
  We are grateful to N. Koval and D. Mu\~noz-Santiburcio for
their help with securing and managing the computational
resources needed for this project. 
  J. F. K. Halliday would like to acknowledge the EPSRC Centre for 
Doctoral Training in Computational Methods for Materials Science 
for funding under grant number EP/L015552/1.
  E. Artacho and M. Famili acknowledge funding from the Leverhulme 
Trust, under Research Project Grant No. RPG-2018-254, 
  E. Artacho is grateful for funding from the EU through the 
ElectronStopping Grant Number 333813, within the 
Marie-Sklodowska-Curie CIG program, and by  the Research 
Executive Agency under the European Union's Horizon 2020 Research 
and Innovation programme (project ESC2RAD, grant agreement no. 776410).
  Funding from Spanish MINECO is also acknowledged, through 
grant FIS2015-64886-C5-1-P, and from Spanish MICIN
through grant PID2019-107338RB- C61/AEI/10.13039/501100011033, 
as well as a Mar\'{\i}a de Maeztu award to Nanogune, Grant 
CEX2020-001038-M funded by MCIN/AEI/ 10.13039/501100011033.
  Computational time from CCSD3, the Cambridge Tier-2 system 
operated by the University of Cambridge Research Computing 
Service funded by the Engineering and Physical Sciences 
Research Council Tier-2 (capital Grant No. EP/P020259/1), 
and DiRAC funding from the Science and Technology Facilities
Council. 
  We also acknowledge the Partnership for Advanced Computing 
in Europe, PRACE, for awarding us access to computational 
resources in Joliot-Curie at GENCI@CEA, France, under 
EU-H2020 Grant No. 2019215186.
\end{acknowledgments}


\appendix*


\section{Review of fundamental approximations}


  The technical approximations for the direct simulation of
electronic stopping processes have been studied before \cite{Kanai2017}.
  Here we review the more fundamental ones underlying the mentioned 
studies and this one.
  As mentioned, they follow the electronic dynamics under the time
dependent external potential originated by the nuclei of both the static
host atoms and the mobile projectile.
  It is done for a finite box in periodic boundary conditions.
  As such it is a well defined time-dependent quantum problem
for the electrons, but its relevance for the real problem relies on
several basic approximations and assumptions, as follows.

\subsection{Non-relativistic quantum mechanics} 

  The described simulations are based on non-relativistic quantum 
mechanics, except for relativistic corrections possibly included in 
the pseudopotentials when containing high-velocity core electrons.
  A relativistic treatment of stopping is known to be needed in the high velocity 
regime, well beyond the Bragg peak \cite{Sigmund}, but the calculations 
discussed here are normally done for projectile velocities up to
and around the Bragg peak.
  For light projectiles the Bragg peak is at around 2 a.u. of velocity.
  Considering that the velocity of an electron out of a collision with
the projectile would be $v_e \lesssim 2 v_p$ (equality for a classical
head-on collision with a heavy projectile), taking $v_e=4$ a.u., 
gives a kinetic energy of $T=217.6$ eV (with a relativistic correction 
$\Delta T = T_{rel} - T_{nr} = 140$ meV). 
  That kinetic energy is smaller than the one of a 2$p$ electron in
an aluminium atom ($\sim 270$ eV), for which relativistic corrections 
are well known to be quite unnecessary for most practical purposes 
of electronic structure.

   The assumption should be taken with care for high-charge 
projectiles, however.
   The Bragg peak goes up to and beyond $v_p \sim 10$ a.u., and an 
 electron with $v_e \sim 20$ a.u. has a kinetic energy of $T\sim 5.5$ keV,
(with $\Delta T \sim 90$ eV), comparable to much deeper core states
for which relativistic corrections are known to be significant.
   Core states, especially of the projectile,  are known to deform 
significantly in the stopping process, away from the free atom 
reference \cite{Ullah2018}, implying that explicitly relativistic 
calculations could be needed when facing the dynamical problem 
including those core electrons.
  It should be kept in mind that in those cases, the core electrons
significantly perturb from their usual atomic state, rendering many
all-electron approaches, such as the augmented plane-wave (APW) 
\cite{madsen2001} or the projector augmented wave (PAW) 
\cite{blochl1994} methods, no suitable for the problem, which 
demands flexible, non-spherical treatment of the relativistic 
Dirac problem.

\subsection{Classical nuclei}

  The approach assumes classical nuclei, which, together with
the quantum electrons define Ehrenfest dynamics.
  It is known that beyond-Ehrenfest approaches are needed 
for the correct long-time thermalisation of the excess energy 
accumulated in the host \cite{Tavernelli-Review,CEID,
abedi2012,gidopoulos2014}.
  Essentially, the spontaneous emission of phonons is a
thermalisation channel that is closed if the quantum fluctuations
of nuclear motion are not included.
  They are much less important for the electronic energy uptake 
from the projectile motion in the short time scale 
\cite{caro2017,caro2019}, 
which is the focus of this and similar electronic stopping studies. 
  Simulating beyond Ehrenfest, including quantum fluctuations for
the nuclear motion, is doable, although considerably more costly 
computationally.
  The most popular approaches are based on the surface hopping 
method and variants thereof \cite{tully1998,SurfaceHoppingReview}, 
but they are generally and better devised for finite systems with well 
separated potential energy surfaces, unlike the continuum of 
excitations encountered in the electronic stopping problem.
  More suitable would be the coupled-electron-ion dynamics approach
of \cite{CEID}, and the many-trajectory sampling with Bohmian 
dynamics of \cite{Tavernelli2013bohmian,tavernelli2020bohmian}, 
although the computational effort is hugely increased.
  Recent progress in exact factorization techniques \cite{abedi2012,
requist2016} also offers interesting possibilities.
  To our knowledge, a quantitative assessment of such effects 
in electronic stopping is still lacking.

\subsection{Constant projectile velocity}  

  In most electronic stopping theory and simulation studies
the projectile is taken to travel at constant velocity.
  It is quite a tradition in the field due to the fact, on one hand, 
that it is inherent to prevalent theories, both in linear response 
\cite{lindhard1954,Reining2016}, 
and in the non-linear theory based on a Galilean transformation 
to the constant-velocity reference frame sitting on the projectile
\cite{Echenique81,Echenique1990,Forcellini2020}.
  On the other hand it is also handy for its applicability in radiation
damage, where the velocity-dependence of the electronic stopping 
power is very generally used.
  It is of course an approximation (see a general assessment in
Ref.~\cite{Correa-Review}), the projectile slowing down 
with a net deceleration of magnitude $a=S_e/m_p$.
  The maximum deceleration would happen for velocities around
the Bragg peak.
 With $S_e\sim 13$ eV/\AA, for example for protons in diamond, 
it amounts to $a=1.4 \cdot 10^{-4}$ a.u. 
  Across the 33.7 Bohr simulation box in the calculations described 
above, within which the steady stopping regime seems to be
well established, an initial velocity of 2 a.u.  would diminish 
by $4.7\cdot 10^{-3}$ a.u., or 0.24~\%.

  Similarly, trajectory deflection is negligible except 
for very small impact parameters.
  This consideration connects with the trajectory sampling 
problem, which has been discussed at length (see e.g.
\cite{Correa-Review} and references therein),
including studies with full Ehrenfest dynamics, which
allow for slowing down and deflection of the projectile. 
  For the purposes of this study, however, a constant velocity 
along high-symmetry channels is considered, as it is a 
problem that is particularly suited for the intended analysis.

\subsection{Electron exchange and correlation} 

  The electron-electron interaction and correlation
in direct electronic stopping simulations is normally 
considered \cite{Correa-Review} within time-dependent 
density-functional theory (TDDFT) \cite{Runge,Marques2006}, 
which is, in principle, exact, except for the approximate 
exchange-correlation (XC) potentials used.
  So far, and to our knowledge, potentials without history 
dependence have been used (local in time), assuming locality 
in space as well (local-density approximation, LDA) or 
generalised gradient approximations (GGA), such
as PBE \cite{PBE}.
  History dependence is included in Ref.~\cite{Nazarov2005}
via linear-response TDDFT in the low-velocity regime.
  There is quite a scope for improvement on this front, both
in the space and time dependence of the XC potential,
although work so far, including validation with experiments,
does not point to XC errors as prominent, within the kind
of simulations studied here.
  The approximation is discussed in some depth for the
electronic stopping problem in \cite{Yost2019,Kanai2021}.
  The work in Refs.~\cite{Nazarov2005,Nazarov2007}
point in the direction of time-dependent current density
functional theory as a very promising route to capture
the non-localities of space and time in a much more
physically meaningful manner.
  Here we still use the adiabatic LDA approximation
(ALDA) and focus on other issues.

\subsection{Conservation} 

  It is important to clarify here some notions on energy
conservation. 
  Starting from the obvious, the dynamics of the projectile itself
is dissipative under the friction provided by the degrees of freedom
of the host, while the dynamics of the whole, projectile plus host
electrons and nuclei conserves the total energy. 
  As mentioned above, the latter conservation is kept within
Ehrenfest dynamics, with quantum electrons and classical 
nuclei.
  Also obviously, the constant-projectile-velocity problem is 
not conservative, the constraining force acting on the projectile
to keep its constant velocity generating a work that increases 
the energy of the host.

  A change of reference frame modifies the energetics. 
  For the particular case of a jellium host, a change of reference
frame was central to the jellium non-linear theory of electronic
stopping \cite{Echenique81}.
  It was changed from the reference in which the electron liquid is
at rest (equilibrium, the laboratory frame) to the one sitting on the 
projectile.
  The Galilean boost transforms the problem of a projectile through 
an homogeneous electron liquid in equilibrium into that of an 
impurity in the homogeneous electron liquid sustaining a
current equal to $-nv_pe$, where $e$ is the charge of the electron,
and $n$ is the constant particle density of the liquid.
  Again the system can be seen as conservative when following 
the dynamics  of all particles, but also as dissipative inasmuch
the electron scattering off the impurity transforms part of the
energy associated to the current into heat.

  It is important to distinguish the above discussion from what
happens in each one of the single-particle scattering events,
or the scattering of each of the auxiliary fermions in the Kohn-Sham
problem in DFT.
  Still in the jellium case, each one of those processes is 
conservative in the projectile reference frame: a quantum particle
scattering off the static external potential defined by the impurity.
  Indeed, the problem has become time-independent, and
it is the time-independent Schr\"odinger equation (or KS) 
the one being solved.

  When changing to the laboratory frame, however, energy
is transferred en each event from the moving projectile 
scatterer to the scattered electron, making it non
conservative.
  This consideration is what gives rise to the stopping power
obtained as derived from Eq.~\eqref{eq:trad-erate}.
  The same discussion is generalised beyond jellium to 
a constant velocity projectile moving along a periodic trajectory
in a crystalline solid. 
  The Kohn-Sham scattering events conserve Floquet 
quasi-energy in the projectile reference frame 
\cite{Forcellini2020}, and a Galilean boost to the laboratory
frame breaks that conservation, in perfect analogy to 
(generalisation of) the jellium case.



\end{document}